\documentstyle[12pt,twoside,fleqn,espcrc1,epsfig]{article}

\newcommand{\AmS}{{\protect\the\textfont2
  A\kern-.1667em\lower.5ex\hbox{M}\kern-.125emS}}

\title{Vector mesons in dense matter }

\author{Su Houng Lee\address{GSI, Planckstr. 1, D-64291 Darmstadt,
 Germany}\address{Department of Physics, Yonsei University, 
Seoul 120-749,    Korea}\thanks{Alexander von Humboldt Fellow.} }
\begin{document}
\maketitle

\begin{abstract}
We will summarize the progress in understanding the changes in the vector meson
spectral density  in nuclear medium using the constraint equations obtained from
the Borel transformed dispersion relation and QCD Operator Product Expansion.  
We will discuss the results for the scalar mass shift and dispersion effects 
(three momentum dependence) for  the light quark system ($\rho, \omega$), the 
strange quark system ($\phi$) and the heavy quark system ($J/\psi$) in nuclear 
medium.  For the light quark systems, a nontrivial change in the mass and width 
are expected, while the dispersion effects are found to be small.   Existing 
model calculations for the dispersion effects are compared to the constraint 
equation in detail.  Very small, but accurate mass shift is obtained for the 
heavy quark system.
\end{abstract}

\section{Introduction}

The properties of vector meson in nuclear medium have been the focus of 
current interest due to their potential role to provide one with a direct 
observable of the  nuclear medium effects, associated with chiral 
symmetry restoration, via dileptons in p-A or A-A reactions\cite{L98}.   
Indeed dileptons from Relativistic Heavy 
Ion Collisions (RHIC)\cite{CERES} seemed to suggest a non-trivial change of 
the 
vector meson spectral density in a hot/dense environment, which can be 
understood in terms of model calculations\cite{LKB95} based on 
decreasing vector meson masses in hot/dense medium\cite{BR91,HL92}.  
However, some model calculations\cite{RCW97,KW98} based on changes of the 
vector meson spectral densities obtained using effective hadronic model
calculations with no decrease in the mass  also seem to 
explain the main features of the CERES data.  
In all of the approaches, the central question is, how the
spectral density changes in hot/dense matter\cite{BR98}.  
In this talk, I will discuss the results for the scalar mass
shift and dispersion effects (three momentum dependence) of the peak position
of the spectral density  for  the light quark system ($\rho, \omega$),
the strange quark system ($\phi$) and the heavy quark system ($J/\psi$)
in nuclear medium.

\subsection{Vector mesons in vacuum}

The distinctive features appearing in the dilepton spectrum in p-A or A-A
reactions are
the vector meson peaks; the $\rho$, $\phi$ and the $J/\psi$, whose
(mass, width) in the vacuum are
(770, 150), (1020, 4.4), (3100, 0.086) MeV.   It is interesting to compare
the phase spaces of their decay into their corresponding  two pseudo 
particles.     For the $\rho$, its two pion decay, which accounts for
most of the total width, has a large phase space.  
This is so because the pion is a  Goldstone boson and has a small mass.  
 For  the  $J/\psi$, its decay
into D-mesons are forbidden, because the D mesons are not Goldstone bosons 
and two times its mass is greater than the mass of  $J/\psi$.   
 For the $\phi$,  the situation is something 
in between  and  
its decay into two kaons has very little phase space.    
Hence, chiral symmetry breaking is partly
responsible for the large difference in their width.
As for the masses of the vector mesons, among other models, QCD sum rules
provide an indirect relation to QCD condensates.   For the light quark system,
 $\langle \bar{q} q \rangle$ is dominantly  responsible for its mass,
for the strange quark system, $\langle \bar{s}s \rangle$ is responsible, 
and for the $J/\psi$, the charmed quark mass and a small non
perturbative contribution from the gluon condensate $\langle
\frac{\alpha_s}{\pi} G^2 \rangle$ are responsible.

Therefore, if chiral symmetry gets restored at finite density or temperature,
non-trivial changes will occur to the masses and widths of the vector
mesons.

\subsection{Quark condensate at finite density}

The temperature dependence of the quark condensates have been calculated long
ago in the lattice\cite{K83,FU86}.  
The result is that the light quark condensate will
go to zero above the critical temperature.  For heavier quark masses,
the changes become smaller.  
At present, due to technical reasons, no lattice result exist at finite
density.  In all what follows, we will use linear density approximation
\begin{eqnarray}
\langle O \rangle_{\rho_n} =\langle O \rangle_0+ \rho_n
\times \langle O \rangle_N
\end{eqnarray}
where $ O$ is any operator, $\rho_n$ the nucleon density, and the subscripts 
$\rho_n, 0 ,N$ denotes
the nuclear, the vacuum and the nucleon expectation values.
Then, we have the following  model
independent result\cite{DL91,CFG92} for the quark condensate
\begin{eqnarray}
\langle \bar{q} q \rangle_{\rho_n}= \langle \bar{q} q \rangle_0
+ { \Sigma_{\pi N} \over 2 \hat{m} } \rho_n  
\sim \langle \bar{q} q \rangle_0 \cdot \biggl(1-0.2 \, \frac{\rho_n}{\rho_0}
\biggr),
\end{eqnarray}
and for the strange part,
\begin{eqnarray}
\langle \bar{s} s \rangle_{\rho_n}= \langle \bar{s} s \rangle_0
+ y { \Sigma_{\pi N} \over 2 \hat{m} } \rho_n  
\sim \langle \bar{s} s \rangle_0 \cdot \biggl(1-0.05 \, \frac{\rho_n}{\rho_0}
\biggr),
\end{eqnarray}
where, $y=2 \langle \bar{s} s \rangle_N /( \langle \bar{u} u \rangle_N
+ \langle \bar{d} d \rangle_N )$ and $\rho_0$ is the nuclear matter 
saturation density.  
One notes that already at nuclear matter, we have partial restoration of
chiral symmetry; namely, the chiral order parameter is reduced 
by 20\%.  Nuclear matter provides a stable environment with non trivial
vacuum changes.  Hence, if anything happens to the vector mesons at high 
temperature or density, the tendencies will already be apparent at nuclear 
matter.

\subsection{Gluon condensate at finite density}

The temperature dependence of the gluon condensates has also 
been calculated on the lattice\cite{CD87,Lee89,AHZ91,KB93}.  
The result is that it will stay almost constant up to the critical temperature
and then reduce to about 60\% of its vacuum value above the critical 
temperature.
The leading density behavior can be obtained from the trace anomaly relation 
to leading order in $\alpha_s$\cite{DL91}
\begin{eqnarray}
T_\mu^\mu=-\frac{9}{8} \frac{\alpha_s}{\pi} G^2+ \sum m_q \bar{q} q
\end{eqnarray}
Taking the nucleon expectation value of it and using the most recent
determination of the nucleon mass in the chiral limit\cite{BM96}, we have
\begin{eqnarray}
\langle \frac{\alpha_s}{\pi} G^2 \rangle_{\rho_n}=
\langle \frac{\alpha_s}{\pi} G^2 \rangle_0 \cdot
\biggl(1-0.05 \, \frac{\rho_n}{\rho_0}
\biggr) 
\label{gluon}
\end{eqnarray}
Hence, we have a non-trivial change in the gluon condensate, although its
relative change is smaller than the case of the quark condensate.

As we have seen, the condensates have  non-trivial change in the  nuclear
medium, which will have a non-trivial effect on the vector meson properties
in nuclear medium.  Experimentally, there are attempts to produce and observe
the decay of vector mesons inside a heavy nuclei and also produce 
meson bound nuclei.
Hence, the theoretical study and the subsequent experimental verification of
changes of vector meson properties in nuclear medium will provide a solid
basis for understanding the hadronic effects in RHIC and medium effects in
general.  

\section{QCD constraints}

There have been many model calculations to study vector meson properties 
in nuclear medium.  Here, we will avoid any model calculation and derive
a constraint equation that any model calculation should satisfy.  
The foundation of this approach was laid in ref.\cite{HL92}

Consider the correlation function of the  vector current 
 $J_\mu=\bar{q} \gamma_\mu q$ at finite density;
\begin{eqnarray}
\Pi_{\mu\nu} (\omega^2, {\bf q}^2 ) &=& i \int d^4x e^{iqx}\langle 
      T [ J_\mu(x) J_\nu(0) ]  \rangle_{\rho_n}.
\label{ope1}
\end{eqnarray}
Here  $q=(\omega,{\bf q})$.
 In what follows, when we give result for explicit vector meson, we will
use the currents  $J_\mu^{\rho,\omega}=\frac{1}{2} ( {\bar u} \gamma_\mu 
 u \mp {\bar d} \gamma_\mu d )$ for the $\rho,\omega$ mesons,
 $J_\mu^{\phi}={\bar s} \gamma_\mu s $ for the $\phi$ and 
 $J_\mu^{J/\psi}={\bar c} \gamma_\mu c $ for the $J/\psi$.

In general, because the vector current is conserved, 
 the polarization tensor in  eq.(\ref{ope1}) will have only two invariant
 functions\cite{tensors}. 
\begin{eqnarray}
\Pi_{\mu\nu}(\omega^2,{\bf q}^2)=\Pi_T q^2 {\rm P}^T_{\mu\nu}+ \Pi_L q^2 
 {\rm P}^L_{\mu\nu},
\label{ope2}
\end{eqnarray}   
where we assume the ground state to be at rest, such that, $
{\rm P}^T_{00} =  {\rm P}^T_{0i}={\rm P}^T_{i0}=0$,  
 ${\rm P}^T_{ij}  =  \delta_{ij}-{\bf q}_i {\bf q}_j/{\bf q}^2 $ and $
{\rm P}^L_{\mu\nu}  =  (q_\mu q_\nu/q^2-g_{\mu\nu}- {\rm P}^T_{\mu\nu})$. 
When ${\bf q} \rightarrow 0$, 
$\Pi_L=\Pi_T$, as in the vacuum.

We will 
 make a small  ${\bf q}$ expansion of the correlation function and  look 
at its  energy dispersion relation at fixed ${\bf q}$,
\begin{eqnarray}
{\rm Re} \Pi_{L,T}(\omega^2,{\bf q}^2)  =  {\rm Re} \left( \Pi^0(\omega^2,0)+ 
\Pi_{L,T}^1(\omega^2,0) ~ {\bf q}^2 + \cdot \cdot \right) \nonumber \\  
  =  \int_0^\infty du^2 \left(
 {\rho^0(u^2,0) \over (u^2-\omega^2)} +  {\rho^1_{L,T}(u^2,0) 
\over (u^2-\omega^2)} ~ {\bf q}^2 + \cdot \cdot \right),
\label{ope3}
\end{eqnarray}
where $\rho(u^2,{\bf q}^2)=1/\pi {\rm Im} \Pi^R(u^2 ,{\bf q}^2)$,  and 
$R$ denotes the retarded correlation function.  
We will construct a constraint equation for $\Pi^0,\Pi_L^1$ and $\Pi_T^1$. 
For $\Pi_L^1,\Pi_T^1$, we will only look at the ``non-trivial'' ${\bf q}$ 
dependence\cite{L98c}.  A simple method to extract the 
``non-trivial'' ${\bf q}$  dependence is to express the polarization function
in terms of $Q^2, {\bf q}^2$ ($\Pi(Q^2, {\bf q}^2)$)
 and extract the linear ${\bf q}^2$ term.  

In general for each polarization functions ($\Pi^0,\Pi_L^1,\Pi_T^1$), 
the OPE \cite{Wilson69,muta} looks as follows,
\begin{eqnarray}
\label{eq2}
\Pi (\omega^2)=\sum_n C_{n} \, 
\langle O_{n} \rangle.
\end{eqnarray}
Here the $O_n$ are operators of (mass) dimension $n$, renormalized at 
a scale $\mu^2$, and $C_{n}$ are the perturbative Wilson 
coefficients, which for the light quark system can be written as  
$C_{n}=\frac{c_n}{(-\omega^2)^{n/2}}$ and for heavy quark system as 
$C_{n}=\frac{c_n}{m_h^n}$.

Let us first discuss the light quark system.  
After the Borel transformation,  the dispersion relation for any one of the
polarization function($\Pi^0,\Pi_L^1,\Pi_T^1$)  becomes
\begin{eqnarray}
{\rm B.T. \,}{\rm Re}\Pi(M^2)=
\int ds \rho(s)e^{-s/M^2},
\label{constraint}
\end{eqnarray}
The left hand side, which is the Borel transform (B.T.) of the OPE,   
looks like
the following when including operators up to  dimension 6, 
\begin{eqnarray}
{\rm B.T. \,}{\rm Re}\Pi(M^2)={\rm B.T. \,}\Pi(M^2)_{pert} 
+\frac{c_4}{1! M^2}\langle O_{4} \rangle 
+\frac{c_6}{2! M^4}\langle O_{6} \rangle 
\end{eqnarray}
The truncation is valid as long as $M^2$ is sufficiently large.  The 
 minimum $M^2_{min}$ is usually determined by requiring the correction from
higher dimensional operators to be less than 30\% of the perturbative 
contribution.
Now the constraints for the spectral density would be eq.(\ref{constraint}), 
applied  above $M^2> M^2_{min}$.  

As can be seen in eq.(\ref{constraint}), the Borel transformation also 
changes the 
weighting factor of the spectral 
density  to an $exp(-s/M^2)$.   This has the following advantage
for practical applications of our constraint.  For small values of the 
Borel mass, the contribution of the spectral density at larger energy is 
exponentially suppressed.  Consequently, in a model calculation, one can 
concentrate on the changes of the spectral density near the vector meson mass
region and below and model the higher energy part with a simple pole like 
contribution.

The constraint equation in the vacuum are well satisfied by the 
spectral density in the vacuum.  As we will see, in most cases the changes 
in the operators $\langle O_{n} \rangle$ are known.  Hence, starting from the 
vacuum form of the spectral density (Im$\Pi$), we can study what  
changes are consistent with the constraint equation.   This provides  
model independent QCD constraints that any model calculation should satisfy.
One can go one step further and try to parameterize the spectral density 
with a simple delta function type of pole and a continuum and determine the 
changes in the  parameters.  

\section{Light quark system $\rho,\omega$}

\subsection{$\Pi_0$}

The operators that dominate the change in the OPE in eq.(\ref{constraint}) 
for the light quark system ($\rho,\omega$) are the quark operators\cite{HL92}
\begin{eqnarray}
\langle O_{4} \rangle & \rightarrow & \langle \bar{q} \gamma_\mu 
D_\nu q \rangle \propto \int dx x [q(x)+\bar{q}(x) ]  \nonumber \\
\langle O_{6} \rangle & \rightarrow & \langle ( \bar{q}  q )^2
\rangle \propto  
\langle ( \bar{q}  q )^2 \rangle_0 + 2 \rho_n \langle \bar{q}  q  \rangle_0
\langle \bar{q}  q  \rangle_N
\label{lpi0}
\end{eqnarray}
The first equation of eq.(\ref{lpi0}) dominates the changes in the dimension 
4 operators and is related to the well known second moment of the quark 
distribution function.  The second equation of eq.(\ref{lpi0}), for which 
we have used the ground state saturation hypothesis\cite{HL92}, 
dominates the changes in the dimension 6 operators.  

Using a delta function assumption for the spectral density in the constraint
equation gives the 
following result for the scalar mass shift at ${\bf q}=0$\cite{HL92},
\begin{eqnarray}
 { m_V(\rho_n) \over m_V(\rho_n=0) }= 1- (0.16 \pm 0.06) {\rho_n \over 
\rho_0}  
\label{lpi0m}
\end{eqnarray}
This result is also consistent with other model 
calculations\cite{Walecka,ST95,FS96} or the Brown-Rho scaling 
argument \cite{BR91}.

A detailed comparison of the constraint equation in eq.(\ref{constraint}) 
with a hadronic calculation, based on chiral SU(3) dynamics with explicit 
vector mesons were performed in \cite{KKW97}.  The result shows a very 
good agreement between the OPE and the phenomenological spectral density 
put into the constraint equation in eq.(\ref{constraint}).   However, 
the hadronic calculation gave a large increase in width with a small decrease 
in mass for the $\rho$ and a large decrease in mass with a small 
increase in width for the $\omega$.  Hence only the result for the   
$\omega$ is consistent
with the result in eq.(\ref{lpi0m}).   Later it was found that this was 
a general result, given the uncertainty in the ground state hypothesis for the
four quark condensate in the medium\cite{LPM98}; 
namely, that there exists a band 
in the mass vs. width plane that satisfies the constraint equation 
in eq.(\ref{constraint}).

\subsection{$\Pi_T$}

 The constraint for the nontrivial ${\bf q}$ dependence in 
eq.(\ref{constraint}) has no $\Pi(M^2)_{pert}$ and has contributions from 
operators with explicit spin index\cite{L98c}.  The contributions from 
dimension 4 operators are related to the twist-2 matrix elements and are well
known.  The dimension 6 operators are dominated also by the twist-2 matrix
elements.   Hence the constraint equation has little uncertainty coming 
from the OPE.  The operators that dominate are 
\begin{eqnarray}
\langle O_{4} \rangle & \rightarrow & \langle \bar{q} \gamma_\mu 
D_\nu q \rangle_N \propto \int dx x [q(x)+\bar{q}(x)]  \nonumber \\
\langle O_{6} \rangle & \rightarrow & \langle \bar{q} \gamma_\mu 
D_\nu D_\alpha D_\beta 
q \rangle_N \propto \int dx x^3 [q(x)+\bar{q}(x) ]
\label{lpit}
\end{eqnarray}

Using a delta function assumption for the spectral density and allowing the 
parameters to change to leading order in density and in ${\bf q}^2$, 
we find the following non-trivial momentum dependence in the peak 
position\cite{L98c},
\begin{eqnarray}
 { m_\rho(\rho_n) \over m_\rho(\rho_n=0) }= 1- (0.023 \pm 0.007) \left(
 { {\bf q} \over 0.5 } \right)^2 {\rho_n \over \rho_0} \nonumber \\ 
 { m_\omega(\rho_n) \over m_\omega(\rho_n=0) }= 1- (0.016 \pm 0.005) \left(
 { {\bf q} \over 0.5 } \right)^2 {\rho_n \over \rho_0},
\label{lpitm}
\end{eqnarray}
where  ${\bf q}$ is in GeV/c unit.  
This shows a very small momentum dependence compared to the expected scalar
mass shift in eq.(\ref{lpi0m}).

A detailed comparison of the constraint equation in eq.(\ref{constraint}) 
for the momentum dependence for the transverse direction has been 
made\cite{FLK99} to the  hadronic calculation, where the  vector-meson nucleon
scattering amplitude is obtained by resonance saturation 
in the  s-channel.  The result shows that the existing model 
calculations tend to overestimate the constraint.  This is due to the 
large   $\rho- N-\Delta(1232)$ coupling,  which is  obtained from the Bonn 
potential\cite{Bonn}.   
However, the existing calculations used a non-covariant 
monopole form factor\cite{RCW97,PPLLM98} normalized off shell
\begin{eqnarray}
 F( {\bf q})= { \Lambda^2 \over \Lambda^2 +{\bf q}^2 }.
\end{eqnarray}
On the other hand, the large $\rho- N-\Delta(1232)$ coupling in the Bonn
potential is defined with a dipole form factor normalized at the on shell
point of the vector meson, 
\begin{eqnarray}
F_\rho(q^2)= \biggl( { \Lambda_\rho^2-m_\rho^2 \over 
\Lambda_\rho^2-  q^2 } \biggl)^{2}.
\label{ff}
\end{eqnarray}
This reduces the Delta contribution to the  rho meson self energy at 
the invariant mass around $m_\Delta-m_N$ by approximately a factor of 4.
After this correction, we find that the model calculations give very 
good agreement with the constraint equation\cite{FLK99}.

\subsection{$\Pi_L$}

The constraint for the longitudinal direction is dominated by 
twist-2 quark and gluon operators 
 $ \langle \bar{q} \gamma_{\mu_1} 
D_{\mu_2} .. D_{\mu_n} q \rangle_N \propto \int dx x^{n-1} 
[q(x)+\bar{q}(x) ],  
 \langle G_{\mu_1}^\alpha 
D_{\mu_2} .. G_{\mu_n \alpha} \rangle_N \propto \int dx x^{n-1}  g(x)$,

Using a delta function assumption for the spectral density and allowing the 
parameters to change to leading order in density and in ${\bf q}^2$, 
we find\cite{L98c},
\begin{eqnarray}
 { m_V(\rho_n) \over m_V(\rho_n=0) }= 1- (0.004 \pm 0.002) \left(
 { {\bf q} \over 0.5 } \right)^2 {\rho_n \over \rho_0},
\label{lpilm}
\end{eqnarray}
 for both the $\rho$ and $\omega$.  This is a very small effect and 
 no detailed comparison with any hadronic calculations exits yet.

\section{Strange quark system $\phi$}

The OPE in the constraint equation for the $\phi$ meson is dominated by 
 $ \langle m_s \bar{s} s \rangle $ for the scalar mass shift and 
 $\langle \bar{s} \gamma_\mu D_\nu s \rangle $ for the momentum dependence. 
Assuming a delta function ansatz for the pole, we find  
\begin{eqnarray}
 { m_\phi(\rho_n) \over m_\phi(\rho_n=0) }= 1- (0.03 \pm 0.015) {\rho_n \over 
\rho_0}  + (0.0005 \pm 0.0002) \left(
 { {\bf q} \over 0.5 } \right)^2 {\rho_n \over \rho_0},
\end{eqnarray}
for the transverse vector meson and the momentum dependence for the 
longitudinal $\phi$ is about a factor of two larger.

\section{Heavy quark system $J/\psi$}

 For the heavy quark system, we look at the constraints from the 
moments\cite{SVZ,RRY},
\begin{eqnarray}
\label{dispersion}
M_n &\equiv& \left. { 1 \over n!} \left( {d \over d \omega^2} \right)^n
{\rm Re} \Pi(\omega^2) \right|_{\omega^2=-Q_0^2} \nonumber \\ &&=
\int_{4m_{c}^2}^{\infty}\frac{\rho(s)}{(s+Q_0^2)^{n+1}}ds.
\end{eqnarray}
The changes in OPE in nuclear medium for $M_n$ is dominated by the 
change in the gluon condensate in eq.(\ref{gluon}).

 To study the $J/\psi$ at rest in the nuclear matter, we will approximate 
the spectral density with a delta function for the lowest state.  This is 
valid  even in nuclear matter, because for a $J/\psi$ at rest, inelastic 
interactions with nucleons such 
as $J/\psi+N \rightarrow \bar{D}+\Lambda_c$ do not occur.
With this assumption, the constraint equation allows for the determination 
of mass shift of the $J/\psi$ in nuclear matter.  We find\cite{KKLMW98} for 
the mass shift
\begin{eqnarray}
\Delta m_{J/\psi} \simeq -7 \, {\rm MeV}.
\end{eqnarray}
This corresponds
to small $J/\psi$- and $\eta_c$-nucleon scattering lengths $a=-\mu_r \Delta m/(2 \pi
\rho_N) \simeq (0.1-0.2)\, $fm ($\mu_r$ is the meson-nucleon reduced mass). 
Our results for the mass shifts of the lowest $\bar{c}c$ states are
surprisingly close to those reported in ref.
\cite{Luke+92,Brodsky+97,Teramond+98,Haya}. 
 
 Although the expected mass shift is small, the result has little uncertainty
coming from OPE and puts reliable
constraint on charmonium mass shift which should be met by further studies of
heavy quark systems in dense matter.

\section{Conclusion}

We have derived and explored the consequences of 
 model independent constraints for the 
vector meson polarization at ${\bf q}=0$ and ${\bf q} \neq 0$ for all the 
vector mesons $\rho,\omega, \phi, J/\psi$.   
Most of them have very little uncertainty in the OPE side of the constraint 
equation and can be used as reliable constraints on all model calculation of 
the vector meson properties in medium.

\section{Acknowledgments}

I would like to thank B. Friman, T. Hatsuda, H. Kim,  S. Kim, F. Klingl, 
P. Morath and W. Weise for the collaboration, which this talk is 
based upon.  This work was supported in part by 
KOSEF through grant no. 971-0204-017-2 and 976-0200-002-2 and the Korean 
Ministry of Education through grant no. 98-015-D00061.

\end{document}